# The Evolution of Sex Chromosomes through the Baldwin Effect

Larry Bull

Computer Science Research Centre

Department of Computer Science & Creative Technologies

University of the West of England, Bristol UK

Larry.Bull@uwe.ac.uk

**Abstract**

It has recently been suggested that the fundamental haploid-diploid cycle of eukaryotic sex exploits a rudimentary form of the Baldwin effect. Thereafter the other associated phenomena can be explained as evolution tuning the amount and frequency of learning experienced by an organism. Using the well-known NK model of fitness landscapes it is here shown that the emergence of sex chromosomes can also be explained under this view of eukaryotic evolution.

# Introduction

The evolution and maintenance of sex is one of the defining characteristics of eukaryotes. Sex is here defined as successive rounds of syngamy and meiosis in a haploid-diploid lifecycle. It has recently been suggested that the emergence of a haploid-diploid cycle enabled the exploitation of a rudimentary form of the Baldwin effect [Baldwin, 1896] and that this provides an underpinning explanation for all the observed forms of sex [Bull, 2017]. The Baldwin effect is here defined as the existence of phenotypic plasticity that enables an organism to exhibit a significantly different (better) fitness than its genome directly represents. Over time, as evolution is guided towards such regions under selection, higher fitness alleles/genomes which rely less upon the phenotypic plasticity can be discovered and become assimilated into the population.

Whether the diploid is formed via endomitosis or syngamy, *the fitness of the cell/organism is a combination of the fitness contributions of the composite haploid genomes*. If the cell subsequently remains diploid and reproduces asexually, there is no scope for a rudimentary Baldwin effect. However, if there is a reversion to haploid cells under meiosis, there is potential for a mismatch between the utility of the haploids compared to that of the diploid; individual haploids do not contain all of the genetic material over which selection operated. That is, the effects of genome combination can be seen as a simple form of phenotypic plasticity for the individual haploid genomes before they revert to a solitary state and hence the Baldwin effect may occur.

In this paper, following [Bull, 2017], the emergence of sex determination chromosomes is explored using versions of the well-known NK model [Kauffman & Levin, 1987] of fitness landscapes where size and ruggedness can be systematically altered. Results suggest that XY (ZW) systems can be shown to be beneficial under certain conditions as they enable further learning – more specifically, fitness landscape smoothing [Hinton & Nowlan, 1987] – to occur on rugged fitness landscapes. The same reasoning also predicts X0 (Z0) systems.

# The NK Model: Asexual Haploid Evolution

Kauffman and Levin [1987] introduced the NK model to allow the systematic study of various aspects of fitness landscapes (see [Kauffman, 1993] for an overview). In the standard model, the features of the fitness landscapes are specified by two parameters: $N$, the length of the genome; and $K$, the number of genes that has an effect on the fitness contribution of each (binary) gene. Thus increasing $K$ with respect to $N$ increases the epistatic linkage, increasing the ruggedness of the fitness landscape. The increase in epistasis increases the number of optima, increases the steepness of their sides, and decreases their correlation. The model assumes all intragenome interactions are so complex that it is only appropriate to assign random values to their effects on fitness. Therefore for each of the possible $K$ interactions a table of $2^{(K+1)}$ fitnesses is created for each gene with all entries in the range 0.0 to 1.0, such that there is one fitness for each combination of traits (Figure 1). The fitness contribution of each gene is found from its table. These fitnesses are then summed and normalized by $N$ to give the selective fitness of the total genome.

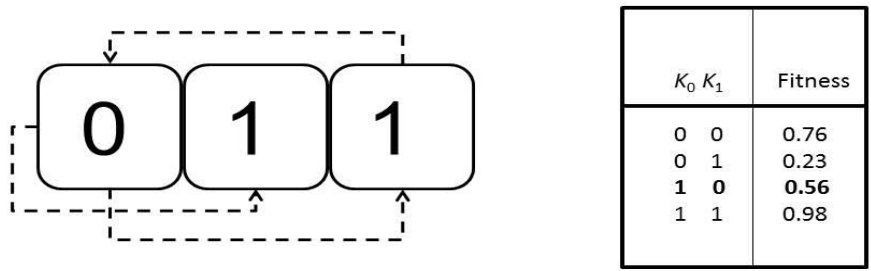

Figure 1: An example NK model ($N$=3, $K$=1) showing how the fitness contribution of each gene depends on $K$ random genes (left). Therefore there are $2^{(K+1)}$ possible allele combinations per gene, each of which is assigned a random fitness. Each gene of the genome has such a table created for it (right, centre gene shown). Total fitness is the normalized sum of these values.

Kauffman [1993] used a mutation-based hill-climbing algorithm, where the single point in the fitness space is said to represent a converged species, to examine the properties and evolutionary dynamics of the NK model. That is, the population is of size one and a species evolves by making a random change to one randomly chosen gene per generation. The "population" is said to move to the genetic configuration of the mutated individual if its fitness is greater than the fitness of the current individual; the rate of supply of mutants is seen as slow compared to the actions of selection. Ties are broken at random. Figure 2 shows example results. All results reported in this paper are the average of 10 runs (random start points) on each of 10 NK functions, ie, 100 runs, for 20,000 generations. Here $0 \leq K \leq 15$, for $N=20$ and $N=100$.

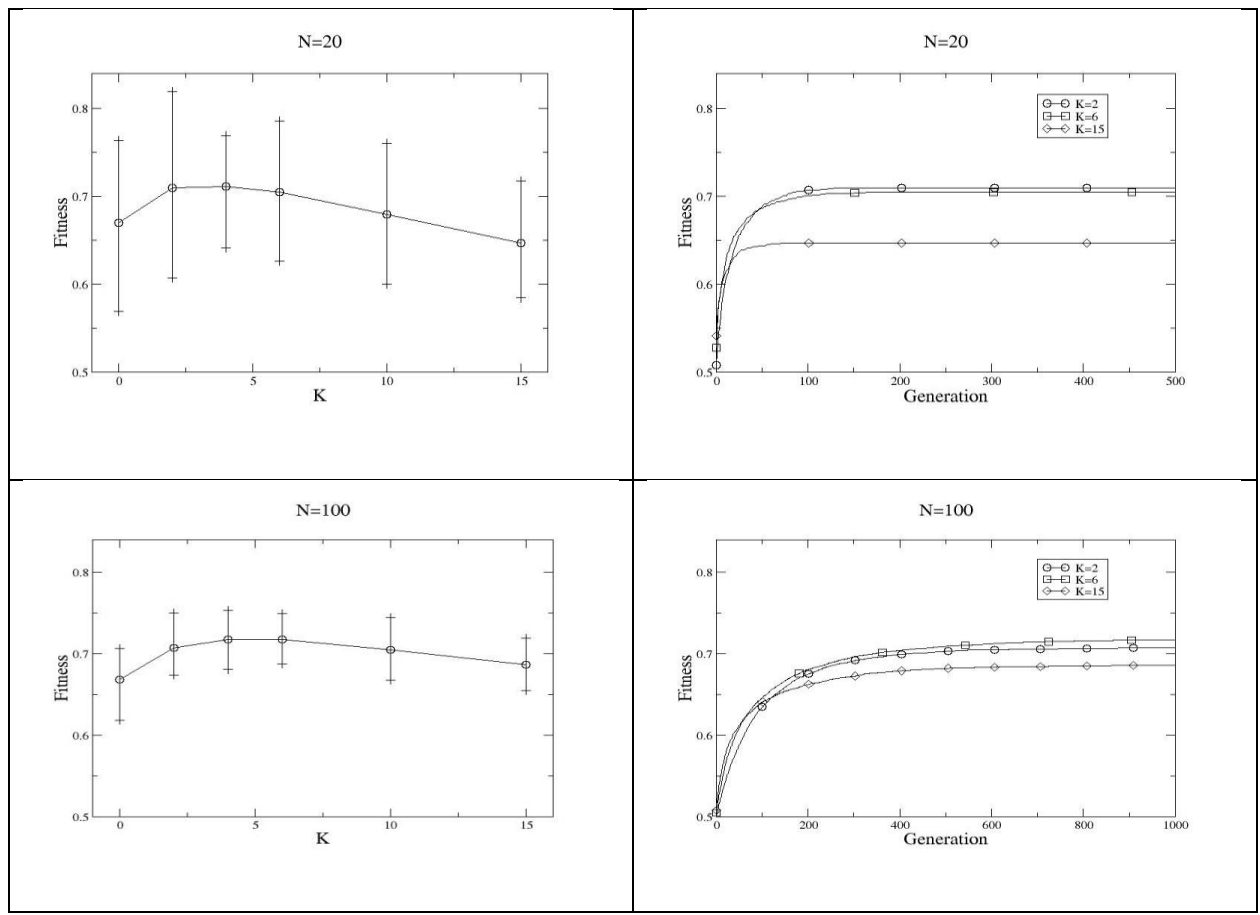

Figure 2: Showing typical behaviour and the fitness reached after 20,000 generations on landscapes of varying ruggedness (K) and length (N). Error bars show min and max values.

Figure 2 shows examples of the general properties of adaptation on such rugged fitness landscapes identified by Kauffman (eg, [1993]), including a "complexity catastrophe" as $K \rightarrow N$. When $K=0$ all genes make an independent contribution to the overall fitness and, since fitness values are drawn at random between 0.0 and 1.0, order statistics show the average value of the fit allele should be 0.66. Hence a single, global optimum exists in the landscape of fitness 0.66, regardless of the value of $N$. At low levels of $K$ (0<$K$<8), the landscape buckles up and becomes more rugged, with an increasing number of peaks at higher fitness levels, regardless of $N$. Thereafter the increasing complexity of constraints between genes means the height of peaks typically found begin to fall as $K$ increases relative to $N$: for large $N$, the central limit theorem suggests reachable optima will have a mean fitness of 0.5 as $K \rightarrow N$. Figure 2 shows how the optima found when $K$>6 are significantly lower for $N$=20 compared to those for $N$=100 (T-test, $p$<0.05).

**Eukaryotes: Sexual Diploid Evolution (Typically)**

Bacteria and archaea can be described as "simpler" than eukaryotes for a variety of reasons including their typically containing a single genome. Hence, as in the NK model above, a given combination of gene values in their genome can be viewed as representing a single point in an $N$-dimensional fitness landscape. In contrast, eukaryotes can contain two or more genomes, typically two, and reproduce sexually via a haploid-diploid cycle with meiosis. Eukaryotes can therefore be viewed as a single point in a fitness landscape of all possible diploids. However, it has recently been suggested that diploid eukaryotes should be viewed as simultaneously representing two points in the fitness landscape of their constituent haploid genomes to explain why sex emerged [Bull, 2017]. Significantly, since the phenotype and hence fitness of a diploid individual is a composite of its two haploid genomes, evolution can be seen to be assigning a single fitness value to the *region* of the landscape between the two points represented by its constituent haploid genomes. That is, it is proposed that evolution is using a more sophisticated scheme by which to navigate the fitness landscapes of eukaryotes than for prokaryotes: an individual organism provides a generalization in the fitness landscape as opposed to information about a single point. This is seen as particularly useful as the complexity of the fitness landscape increases in both size and ruggedness.

Moreover, the haploid-diploid cycle can also be explained as creating a rudimentary form of the Baldwin effect [Baldwin, 1896], thereby enabling the beneficial smoothing of rugged fitness landscapes [Hinton & Nowlan, 1987]. Briefly, under the Baldwin effect, the existence of phenotypic plasticity potentially enables an organism to display a different (better) fitness than its genome directly represents. Typically, such plasticity is assumed to come from the organism itself, eg, through the modification of neural connectivity. However, a genetically defined phenotype can also be altered by the exploitation of something in the environment, eg, a tool. It is suggested that haploids forming pairs to become diploid is akin to the latter case of learning/plasticity with the partner's genome providing the variation. Being diploid can potentially alter gene expression in comparison to being haploid and hence affect the expected phenotype of each constituent haploid alone since both genomes are active in the cell - through changes in gene product concentrations, partial or co-dominance, etc. If the cell/organism subsequently remains diploid and reproduces asexually, there is no scope for a rudimentary Baldwin effect. However, if there is a reversion to a haploid state for reproduction under a haploid-diploid cycle, there is the potential for a significant mismatch between the utility of the haploids passed on compared to that of the diploid selected; individual haploid gametes do not contain all of the genetic material through which their fitness was determined.

The variation processes can then be seen to change the bounds for sampling combined genomes within the diploid by altering the distance between the two end points in the underlying haploid fitness landscape. That is, the degree of possible change in the distance between the two haploid genomes controls the amount of learning possible per cycle. Numerous explanations exist for the benefits of recombination (eg, [Bernstein and Bernstein, 2010]) but the role becomes clear under the new view: recombination moves the current end points in the underlying haploid fitness space which define the fitness level generalization either closer together or further apart. That is, recombination adjusts the size of an area assigned a single fitness value, potentially enabling higher fitness regions to be more accurately identified over time. Moreover, recombination can also be seen to facilitate genetic assimilation within the simple form of the Baldwin effect: if the haploid pairing is beneficial and the diploid cell/organism is chosen under selection to reproduce, the recombination process can bring an assortment of those partnered genes together into new haploid genomes. In this way the fitter allele values from the pair of partnered haploids may come to exist within individual haploids more quickly than the under mutation alone (see [Bull, 2017] for full details). Mutation can also be seen to be adjusting the distance between the two genomes at a generally reduced rate per generation. It has previously been shown how both the amount of learning per step and the rate at which it occurs can affect utility, with more learning typically proving increasingly beneficial with increasing $K$ [Bull, 1999].

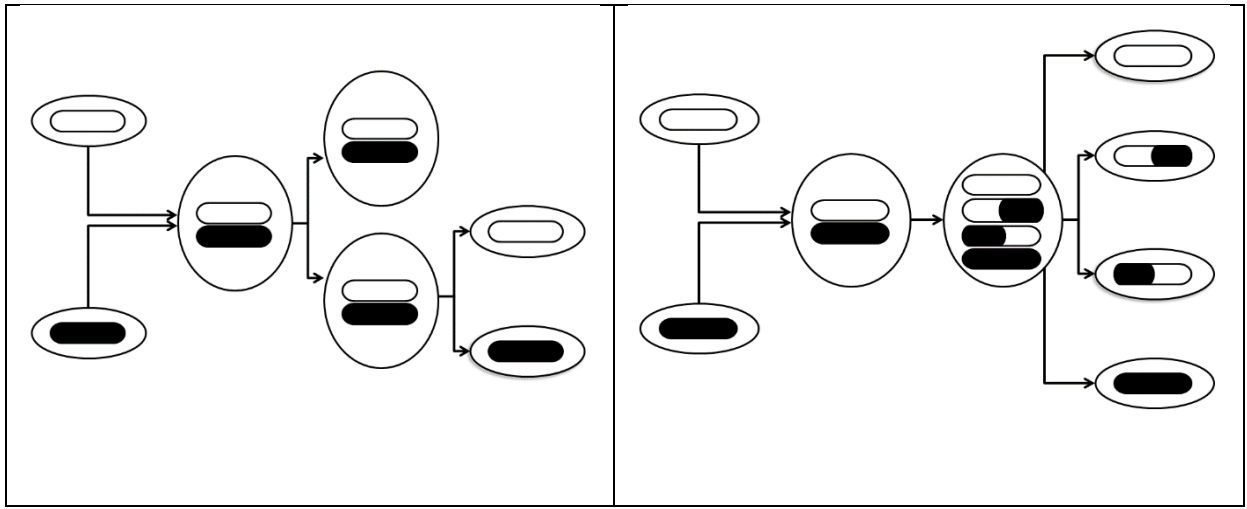

Figure 3: The one-step meiosis (left) and two-step meiosis with recombination (right) processes explored here (after [Maynard Smith & Szathmary, 1995, p151]).

**One-step Meiosis and Two-step Meiosis with Recombination in the NK Model**

As discussed in [Maynard-Smith & Szathmary, 1995, p150] the first step in the evolution of eukaryotic sex was the emergence of a haploid-diploid cycle, probably via endomitosis, then simple syngamy or one-step meiosis, before two-meiosis with recombination (Figure 3). Following [Bull, 2017], the NK model can be extended to consider aspects of the evolution of sexual diploids. Firstly, each individual contains two haploid genomes of the form described above for the standard model. The fitness of an individual is here simply assigned as the average of the fitness of each of its constituent haploids. These are initially created at random, as before. Secondly, simple syngamy is here implemented as follows: on each generation the diploid individual representing the converged population is copied twice to create two offspring. One of the two haploids in each offspring individual is chosen at random.

Finally, a random gene in each chosen haploid is mutated. The resulting pair of haploids forms the new diploid offspring to be evaluated.

Two-step meiosis with recombination is here implemented as follows: on each generation the diploid individual representing the converged population is similarly copied twice to create two offspring. In each offspring, each haploid genome is copied once, a single recombination point is chosen at random, and non-sister haploids are recombined. One of the four resulting haploids in each offspring individual is chosen at random. Finally, a random gene in each chosen haploid is mutated. The resulting pair of haploids forms the new diploid offspring to be evaluated.

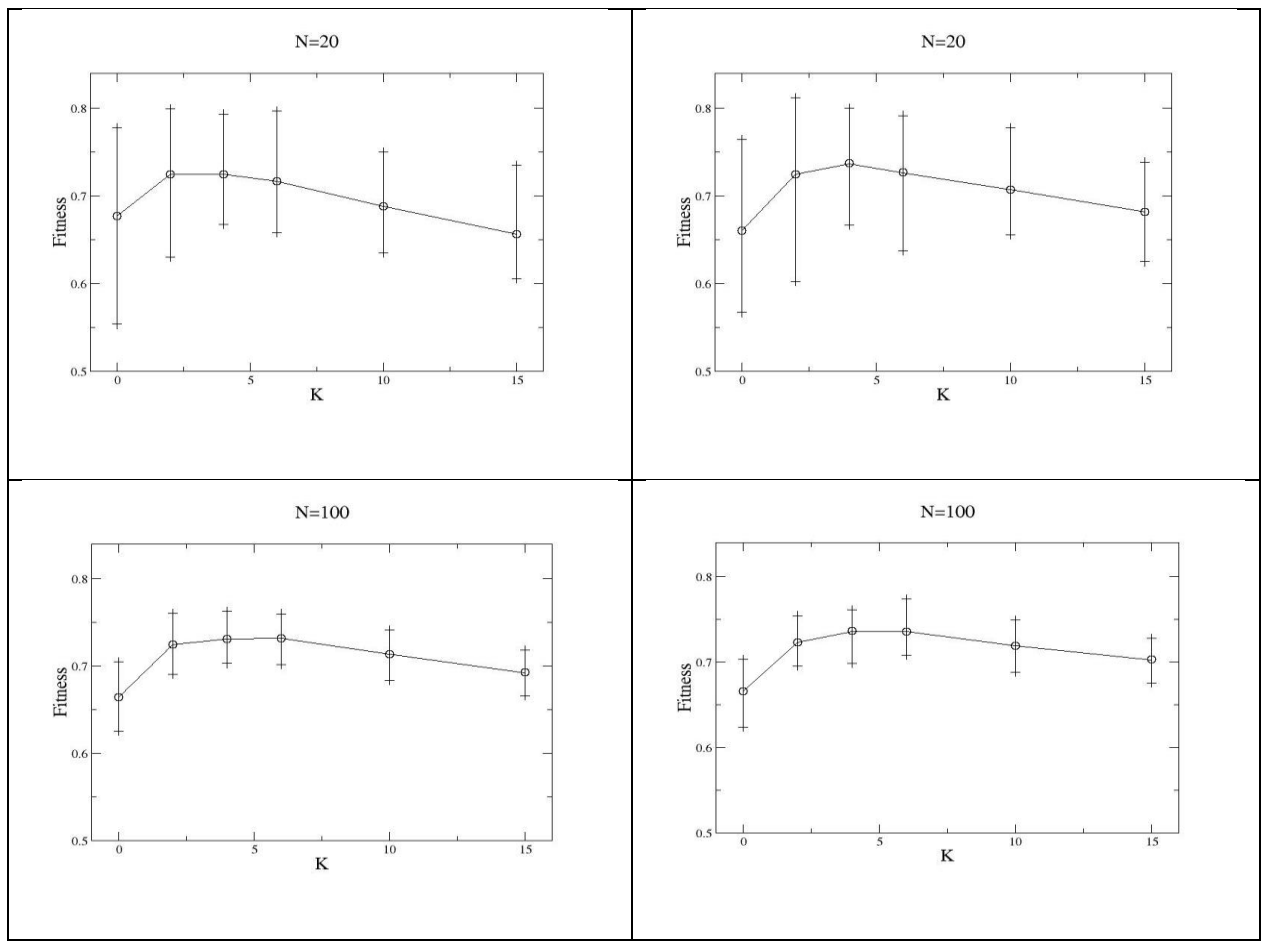

Figure 4: Showing typical behaviour and fitness reached after 20,000 generations on landscapes of varying ruggedness (*K*) and size (*N*) for diploids undergoing one-step meiosis (left column) or two-step meiosis with recombination (right column).

Figure 4 shows examples of the general properties of adaptation in the NK model of rugged fitness landscapes for diploid organisms evolving via either simple one-step meiosis or two-step meiosis with recombination. Under both forms of meiosis, when $N$=20 the fitness level reached is significantly lower than for $N$=100 for $K$>4 (T-test, $p$<0.05), as was seen in the traditional haploid case due to the effects of the increased landscape complexity. Following [Bull, 2017], it can be seen that fitness levels are always higher than the equivalent haploid case (Figure 2) when $K$>0 due to the Baldwin effect as discussed (T-test, $p$<0.05). Again, as reported in [Bull, 2017], the fitness levels are always higher for two-step meiosis with recombination in comparison to simple syngamy for $K$>2 (T-test, $p$<0.05). This is explained by the potential for an increased rate of change in the distance between the two genomes and hence the effective amount of learning experienced per generation (after [Bull, 1999]). Figure 5 shows examples of how the size of those generalizations changes over time depending upon the ruggedness of the landscape and the form of meiosis; assimilation through recombination can be seen.

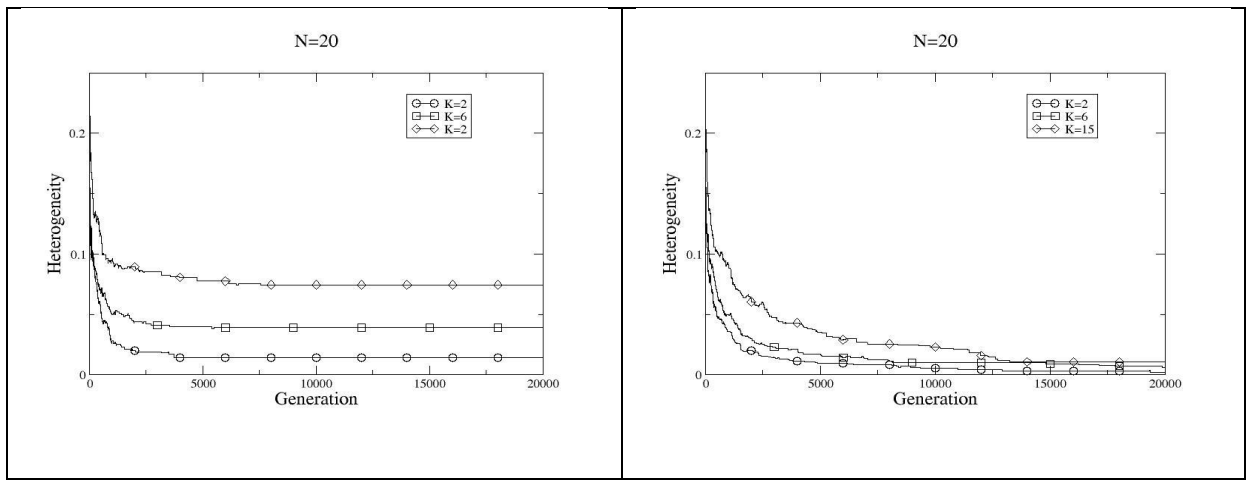

Figure 5: Showing typical convergence behaviour as a fraction of the difference in corresponding gene values in the two haploid genomes on landscapes of varying ruggedness ($K$) for diploids undergoing one-step meiosis (left) and two-step meiosis with recombination (right).

**Sex Chromosomes**

The emergence of isogamy, ie, mating types, was not considered in the explanation for the evolution of two-step meiosis with recombination above (as in [Bull, 2017]). However, the presence of allosomes - XY in animals and ZW in birds, some fish, reptiles, insects, etc. – can also be explained as a mechanism by which a haploid genome may vary the amount of learning it experiences when paired with another to form a diploid organism. Importantly, taking the view of the constituent haploid genomes, the presence of an heterogametic sex creates the situation where, as evolution converges upon optima, a given haploid containing the common (X or Z) allosome will typically experience two different fitness values simultaneously within a population due to genetic differences between the two sexes; two fitness contributions from the common allosome will almost always exist with two mating types. *It is here proposed that the extra (approximate) fitness value information can prove beneficial to the learning/generalisation process described above by adding further landscape smoothing.*

To introduce autosomes and allosomes to the above diploid model, the original pair of haploid genomes of length *N* are each subdivided into *n*=2 equally sized chromosomes, ie, there are 2*n* chromosomes per diploid. A (converged) sub-population of a homogametic sex is said to exist along with a (converged) sub-population of a heterogametic sex. No functional differentiation is imposed upon the heterogametic sex fitness function here; the fitness landscapes of both sexes are identical.

Autosomes undergo two-step meiosis with recombination, as above, whereas allosomes do not undergo recombination. The sex of the offspring is determined by which allosome is (randomly) selected from the heterogametic sex. Once the resulting overall diploid genome is created, mutation is applied to each haploid as before. The fitness contribution of the haploid genomes is their average, as above. For example, when X-inactivation occurs in mammals the choice is typically random per cell lineage in the placenta and hence the fitness contribution of the allosomes remains a composite of the two chromosomes.

Figure 6 shows examples of how the benefits of sex chromosomes can vary with landscape ruggedness and size. Note the average fitness of the heterogametic and homogametic sexes is shown. An equivalent (converged) population of hermaphrodites is used for comparison here, where recombination is not used for the second chromosome. As can be seen, for *N*=20, two sexes prove beneficial for *K*>4 (T-test, *p*<0.05) over the hermaphrodite. In contrast, with *N*=100, two sexes prove detrimental under the same conditions (T-test, *p*<0.05). Therefore the improvement in fitness seen for sex cannot simply be attributed to the presence of two explicit sub-populations. There is no significant difference in fitness seen for $K<6$ (T-test, $p≥0.05$) for either $N$. It can also be noted that including recombination in the second chromosomes (allosomes) of the hermaphrodites does not significantly affect their fitness under any conditions explored (not shown, T-test, $p≥0.05$).

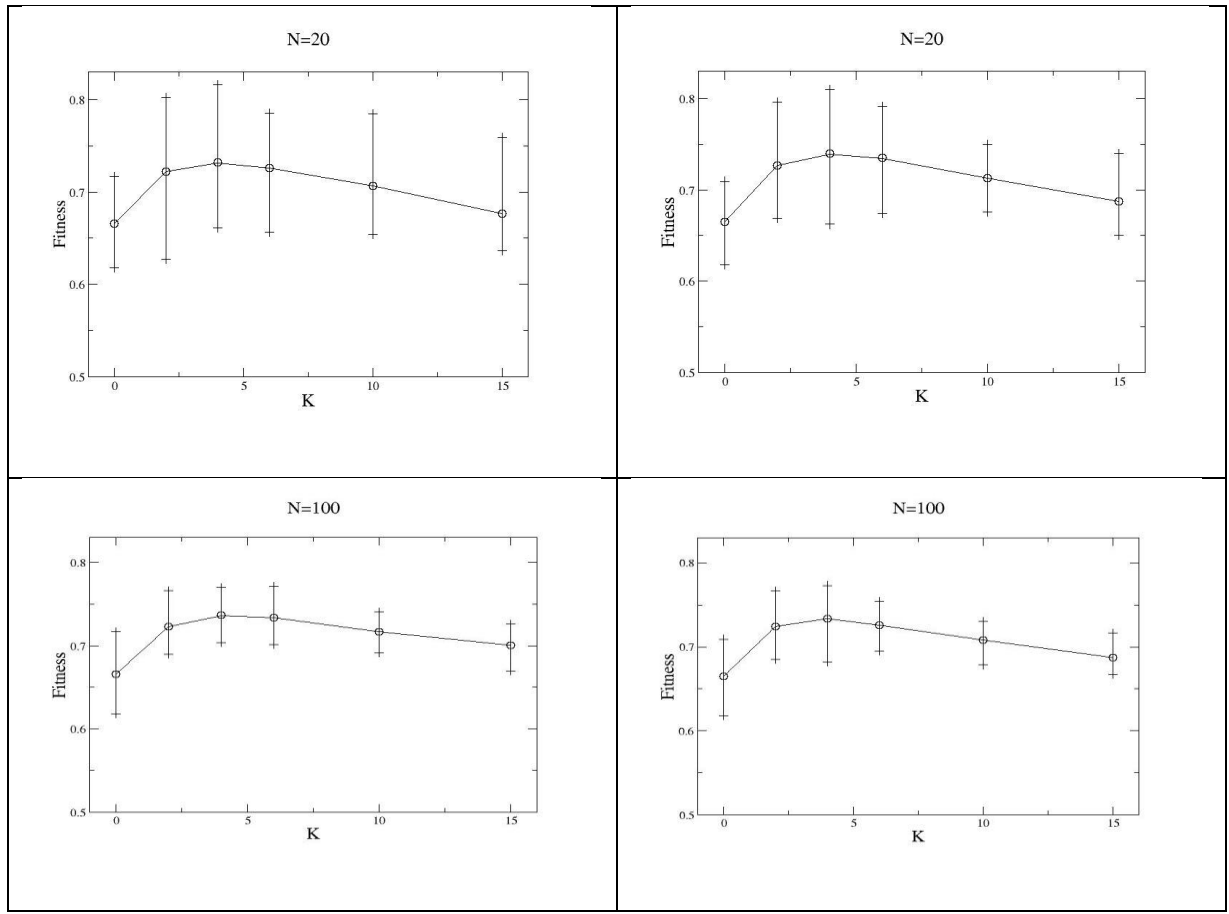

Figure 6: Showing typical behaviour and fitness reached after 20,000 generations on landscapes of varying ruggedness ($K$) and size ($N$) for hermaphrodite diploids (left column) or with two sexes (right column).

As discussed in [Bull, 2017], increasing the amount of learning under a standard Baldwin-effect scheme gives increasing benefit for $K>6$ with $N=20$, whereas increasing the amount of learning decreases fitness for $K>4$ with $N=100$. As noted above, it is here suggested that the presence of sex chromosomes creates another mechanism through which the fitness landscape of the constituent haploids with the common allosome is potentially smoothed. That is, as the degree of heterogeneity between the two haploids in the homogenetic sex converges (Figure 5, right) two different fitness values are maintained due to the fitness obtained within the heterogametic sex. This is in contrast to hermaphrodites where eventual near convergence of the two haploids reduces the amount of learning/smoothing as an optimum is found. Therefore the results in [Bull, 2017] predict those seen here with sex chromosomes added almost exactly: the extra learning mechanism is useful for higher $K$ when $N=20$ and detrimental for higher $K$ when $N=100$. Although the extra smoothing proves beneficial on less rugged landscapes ($K>4$) than in the standard case with $N=20$ perhaps due to the convergence of the two haploids over time reducing the amount of learning experienced from their partnering.

It can be noted that, whilst varying between reproducing with a member of the opposite sex and as a hermaphrodite is seemingly relatively common in nature (eg, see [Bachtrog et al., 2014]), no significant change in behaviour was seen here for a variety of ratios for the parameters explored (not shown). However, the explanation presented that either form of reproduction provides a difference in the amount of learning exhibited suggests a selective advantage for species able to suitably tune the ratio between the two based upon current conditions.

**Discussion**

The X0 (Z0) system also enables a variation in the amount of learning, typically reduced from XY (ZW) since one mating type will provide the "pure" fitness of a single X (Z) chromosome. Indeed, results (not shown) from a version of the above model containing X0 finds no significant change in fitness over XY except when $N=20$ and $K>6$ (T-test, $p<0.05$). A general benefit of the XY (ZW) system over X0 (Z0) exists when the diploid fitness landscape is considered. As shown in Figure 5 (right), hermaphrodites exploiting two-step meiosis with recombination can be expected to eventually converge upon organisms carrying two copies of the same (or very nearly) haploid genome. This means that over evolutionary time they become increasingly restricted to the region of symmetry within the diploid fitness landscape, ie, the region where constituent haploid genome A is genetically similar to haploid genome B. However, in the presence of two mating types, high levels of fitness may exist in other areas of the diploid landscape, as shown in Figure 7. One way to avoid a near complete set of homozygotes is to maintain a region(s) in the haploid genome where recombination does not occur, thereby inhibiting the assimilation process. It is here suggested this is what the XY (ZW) system enables over the X0 (Z0) system. Asexual reproduction and one-step meiosis, ie, a haploid-diploid cycle without recombination, can also hinder convergence (Figure 5, left). Whilst the results in [Bull, 2017] suggest two-step meiosis with recombination is generally beneficial over asexuality and one-step meiosis due to the increased amount of learning, there is some overlap at low levels of landscape ruggedness. Parabasalid are known to exploit simple syngamy and since they have lost their mitochondria may be seen to exist on less rugged landscapes (after [Bull & Fogarty, 1996]). It can be speculated that they may therefore be exploiting the potential to move away from the region of symmetry within their diploid fitness landscapes. Having adopted the generally more beneficial two-step meiosis with recombination, the XY (ZW) system enables eukaryotes to maintain heterozygotes for some regions of their overall genome space in a relatively controlled manner. This becomes particularly significant if the presence of mating types also causes an increase in fitness in the region of symmetry, eg, through dimorphism.

Figure 8 shows results from introducing such heterogeneity/difference into the fitness landscape of the two mating types used above. Here the fitness values in the table for the X chromosome are constructed as usual and those for the Y chromosome are made by subtracting the corresponding value from 1.0, ie, so that the opposite gene values are preferred. As expected, performance is typically improved when recombination between X and Y chromosomes is not allowed but is between X chromosomes ($K>4$, T-test $p<0.05$) when $N=100$. Results with $N=20$ do not show any significant difference in fitnesses reached (not shown). The equivalent hermaphrodite is always worse, regardless of $N$ and $K$ (not shown).

Hence this explains such things as why recombination does not typically occur between allosomes but why recombination over some regions is sometimes seen; the degree of difference is controllable. Relatedly, the two sex chromosomes can be of different sizes which is also potentially correlated to the degree of heterogeneity required between the two haploids to reach the higher fitness areas. Note the situation can be reversed from the XY system in the ZW system since which mating type maintains the single copy of the common allosome is not important in exploiting the benefits described. It also helps to explain why more than two mating types exist in some species - the presence of more than one high fitness region away from the area of symmetry in the fitness landscape is exploitable by the maintenance of a corresponding mating type per region. As well as serving as a mechanism through which the underlying haploid landscape is smoothed through "appropriate" genome pairings, as noted in [Bull, 2017], sexual selection can also be seen to aid the identification of the higher optima away from the region of symmetry when many exist. Higher optima existing in the region of symmetry through the existence of two mating types would explain the maintenance of sex in even the most unchanging environments. Conversely, that the existence of such regions may vary temporally could help to explain environmental sex determination mechanisms. Moreover, similar reasoning would seem to apply regardless of the details of the mechanism, eg, for polygenic sex, cytoplasmic control, etc.

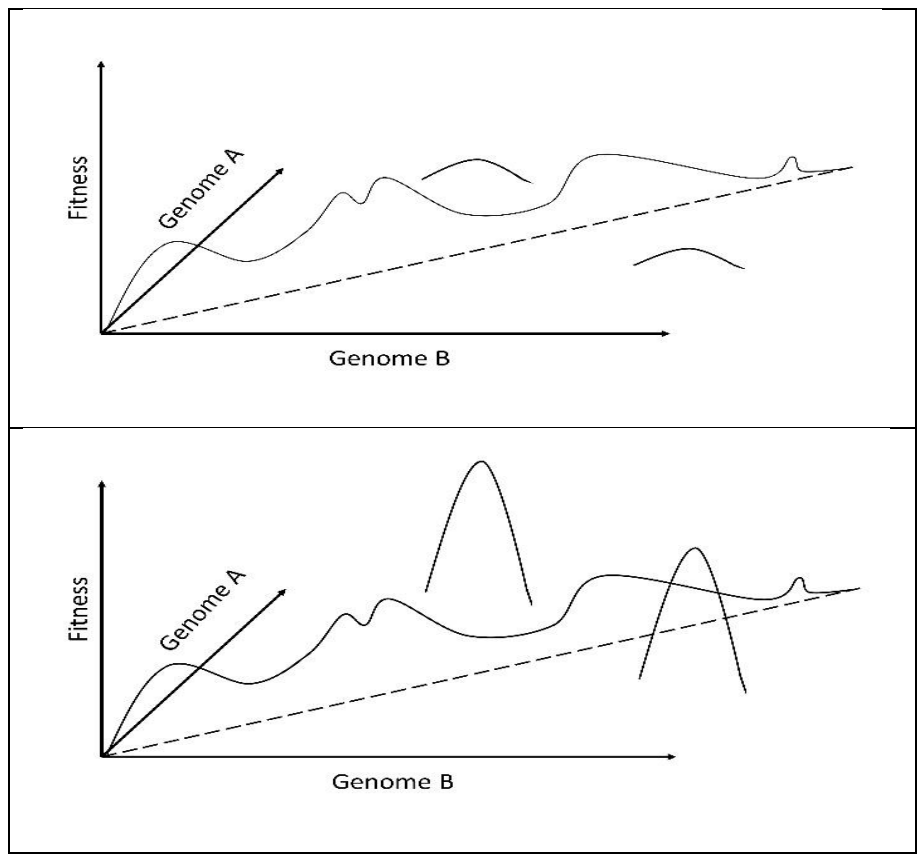

Figure 7: Showing simple example fitness landscapes where mating types are not (top) and are expected (bottom) to prove beneficial. In the second case, high optima also exist away from the area of genome symmetry within the diploid landscape. The existence of sex chromosomes enables the males (females) to maintain genomes corresponding to the end-point on the axis of possible gene combinations and therefore occupy the outlier optima.

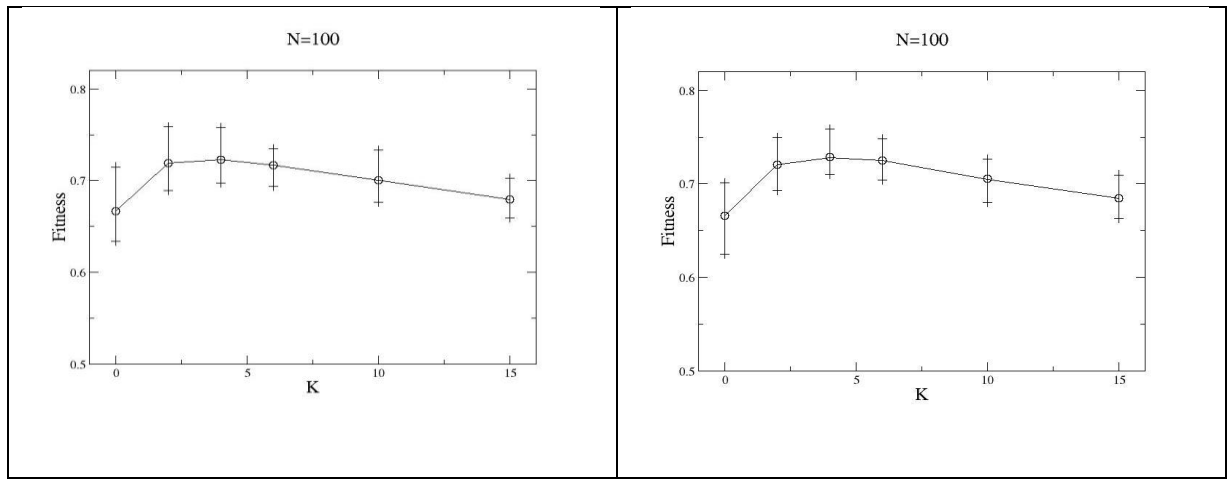

Figure 8: Showing typical behaviour and fitness reached after 20,000 generations on landscapes of varying ruggedness ($K$) with a heterogeneous component in the sex chromosome, with (left) and without (right) recombination between X and Y chromosomes.

**Conclusion**

The evolution of allosomes was explored and a beneficial fitness landscape smoothing effect from the extra (approximate) fitness value information was found for smaller, more rugged landscapes. Since no functional differentiation or recombination was used for the allosomes, the results here apply equally well to the emergence of XY and ZW systems. Moreover, the same reasoning above applies equally well to the emergence of X0 and Z0 systems, although the effect from the extra fitness value would potentially reduce over time as the homogametic sex converges depending upon any dosage effects, etc. Finally, the role – or not – of recombination for allosomes was explored when explicit mating type fitness difference is assumed, clarifying previous discussions [Bull, 2020, p59-60].

The results and discussions here reinforce the basic hypothesis presented in [Bull, 2017]: eukaryotic sex exploits a rudimentary form of the Baldwin effect.